%% file: paper.tex
\documentclass[twocolumn,showpacs,aps,prd,superscriptaddress]{revtex4}
\usepackage{graphicx,rotating}
\usepackage{dcolumn}
\usepackage{epsfig}
\usepackage{multirow}
\input{pubboard/babarsym.tex}

\def\Bu{\mbox{$\mathrm{B_u}$}}

\def\Dstar{\mbox{$\mathrm{ D^{\star}}$}}
\def\Dz{\mbox{$\mathrm{ D^0}$}}

\def\gevcsq{\mbox{GeV/$c^2$}}

\long\def\inst#1{\par\nobreak\kern 4pt\nobreak
    {\it #1}\par\vskip 10pt plus 3pt minus 3pt}
\begin{document}

\begin{flushleft}
BABAR-PUB-05/046\\
SLAC-PUB-11619 \\
hep-ex/0601017 \\
\end{flushleft}

 \title{{\boldmath Measurements of $\Lambda^+_c$ Branching Fractions of Cabibbo-Suppressed Decay Modes involving $\Lambda$ and $\Sigma^{0}$.}}

\input pubboard/authors_sep2005.tex
\date{September 19, 2006}

\begin{abstract}
\input{abstract.tex}

\end{abstract}

\maketitle
\setcounter{footnote}{11}

\section{Introduction}
A considerable body of work has been done on charmed baryons~\cite{PDG}. Nonetheless,
%Beginning with the first observation of the charmed baryon $\Lambda_c^+$ in the1970s by BNL and MARK-II at SLAC~\cite{Samios, MARK},
our understanding of the physics of charmed baryons has developed less rapidly than that of charmed mesons. This is due to the smaller baryon production cross sections, shorter lifetime, and, in \epem storage rings, the absence of cleanly observable
%high signal to background
 $\Lambda_c$ $\overline{\Lambda}_c$ resonances. During the last few years there has been significant progress in the
experimental study of the hadronic decays of charmed baryons. Recent results
 on masses, widths, lifetimes, production rates, and the decay asymmetry parameters of the charmed baryons have been
published by different experiments; among them are the observations of Cabibbo-suppressed decays $\Lambda^+_c \to \p\phiz$ by the CLEO collaboration~\cite{CLEO}, and $\Lambda^+_c \to \Lambda K^{+}$ and $\Lambda^+_c \to \Sigma^{0}K^{+}$ by the Belle collaboration ~\cite{Belle}.

The precision in the measurements of branching fractions is only about
 40$\%$ for many Cabibbo-favored charm baryon modes~\cite{PDG}, while for 
Cabibbo-suppressed decays the precision is even
worse. As a consequence, we are not yet able to distinguish between
the decay rate predictions made by different models, e.g., the quark model approach to non-leptonic charm decays~\cite{Khana,Kram} and Heavy Quark Effective Theory~\cite{Kron}. Only one model~\cite{Khana} gives predictions for the Cabibbo-suppressed decays.

In this paper we present a study of $\Lambda_c^+$ baryons produced in \epem $\to$ \qqbar ($q = u, d, s$ or $c$ quark) interactions at \babar. We present improved measurements of the Cabibbo-suppressed decays $\Lambda^+_c$ $\to$  $\Lambda K^+$ and $\Lambda^+_c$ $\to$ $\Sigma^{0} K^+$ relative to Cabibbo-favored decays $\Lambda^+_c$ $\to$ \lz\ $\pi^+$ and $\Lambda^+_c$ $\to$ $\Sigma^{0}$ $\pi^+$, respectively, and set an upper limit on the decay modes $\Lambda^+_c$ $\to$ $\Lambda K^+ \pi^+ \pi^-$, and $\Lambda^+_c$ $\to$ $\Sigma^{0} K^+ \pi^+ \pi^-$ relative to the same Cabibbo-favored decays. Here and throughout this paper, inclusion of charge-conjugate states is implied.
\vspace{-0.1in}
\section{The \babar\ detector and data samples}
\label{detector}
The data used in this analysis were collected by the \babar\ detector at the \pep2
storage ring. We use data corresponding to a total integrated luminosity of $112~\mathrm{fb}^{-1}$ taken at the \FourS resonance (on-resonance) 
and $13~\mathrm{fb}^{-1}$ taken at a center-of-mass (CM) energy
$40\mev$ below the \FourS mass (and below the threshold of \BB production, the off-resonance). 
A detailed description of the \babar\ detector is presented in
Ref.~\cite{ref:babar}. The components of the detector most
relevant to this analysis are described here. Charged-particle tracks are reconstructed with a five-layer, double-sided
silicon vertex tracker (SVT) and a 40-layer drift chamber (DCH) with a
helium-based gas mixture, placed in a 1.5-T solenoidal field produced
by a superconducting magnet. The resolution in $p_T$, the track momentum transverse to the beam direction, is approximately
$(\delta p_T/p_T)^2 = (0.0013\ (\mbox{GeV}/c)^{-1}\, p_T)^2 + (0.0045)^2$.
Kaons and protons are identified with likelihood ratios calculated from the ionization energy loss
(\dedx) measurements in the SVT and DCH, and from the observed pattern of Cherenkov light in an internally reflecting ring imaging detector (DIRC). The efficiency for
identifying true kaons exceeds 80$\%$, while the probability for a pion to be misidentified as a
kaon is less than 3$\%$. Photons are identified as
 isolated electromagnetic showers in a CsI(Tl) electromagnetic calorimeter. Large Monte Carlo (MC) samples generated with JETSET \cite{jetset} are used to determine signal detection efficiency. The detector response in these samples is simulated with the GEANT4 \cite{G4} program. Particle identification efficiencies are corrected using data control samples. 

\section{Event and candidate selection}\label{sec:Analysis}

 \indent Candidates for $\Lambda$, which is in the final state of all the decay modes involved in this analysis, are reconstructed in the decay mode $\Lambda \to p \pi^{-}$.
 We fit the $p$ and $\pi^{-}$ tracks to a common vertex and require the $\chi^{2}$ probability of the vertex fit to be greater than 0.1$\%$. The three-dimensional flight distance of each $\Lambda$ candidate between its decay vertex and the event primary vertex is required to be greater than 0.2~cm. 

This analysis is based on fits of invariant masses or differences between invariant masses, in the case of $\Sigma^0 \ra \Lambda \gamma$ decays.
%%mostly for the decays which involve $\Sigma^0$ ($\Sigma^0 \ra \Lambda \gamma$), as well as for the decays where we do not have statistically significant signal. 
%this procedure is used to improve the mass resolution significantly. 
In general, the fits are performed with the following criteria. For the signal, the sum of two Gaussian functions with a common mean (the two widths, the common mean and the fraction of the core Gaussian being free as discussed in detail in the corresponding fits below) is the preferred function to better reproduce the tails. However for the decay modes with less statistics, a single Gaussian with free mean and width is used. In case there is no statistically significant signal, a single Gaussian with fixed mean and width has to be used. The background is parametrized by a polynomial in invariant mass with order 2, or higher, as required to obtain a satisfactory fit.
%As for the background, the basic function is a $2^{\mathrm{nd}}$ order polynomial. When it does not do a good enough job, based on the $\chi^2$ per degree of freedom for the fit, a $3^{\mathrm{rd}}$ (or even higher) order polynomial is used.
 
\begin{figure}
\begin{center}
\includegraphics[width=3.5in]{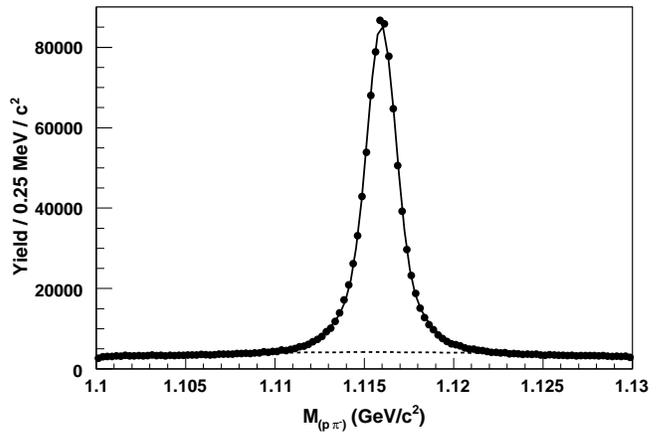}
\caption{The invariant mass of $p \pi^{-}$ combinations. The solid line indicates the result of the fit for the sum of the signal and background and the dashed line for the background only.}
\label{fig:lambda}
\end{center}
\end{figure}

The invariant mass of $\Lambda$ candidates is fitted using a sum of two Gaussian functions with a common mean to represent the signal and a 2$^{\mathrm{nd}}$ order polynomial to represent the background. The fitted distribution is shown in Fig.~\ref{fig:lambda}. The resolution is measured to be $\sigma_{\rm {RMS}} = 1.5~\mevcsq$, where $\sigma_{\rm {RMS}}$ is defined by
\[ \sigma_{\rm {RMS}}^{2} \equiv f_{1} \sigma_{1}^{2} + f_{2} \sigma_{2}^{2},\]
where $\sigma_{1} = 0.820 \pm 0.003~\mevcsq$ (the width of the core Gaussian) and $\sigma_{2} = 2.103 \pm 0.021~\mevcsq$ (the width of the wider Gaussian), and $f_{1}$ and $f_{2}$~(= $1 - f_{1}$) are the two corresponding fractions of the two Gaussian functions, with $f_{2}$ = 42\% of candidates in the wider Gaussian. The mass of a $\Lambda$ candidate, used in the reconstruction of $\Lambda_c^+$ or $\Sigma^{0}$ decays, is required to be in the range 1113~\mevcsq~$<$ $M_{p\pi^{-}}$ $<$ 1119~\mevcsq.
%where $f_{1}$ and $f_{2}$ are the fractions of the two Gaussian functions, and $\sigma_{1}$ and $\sigma_{2}$ are the two corresponding widths. 
%The mass of a $\Lambda$ candidate is required to be in the range 1113~\mevcsq~$<$ $M_{p\pi^{-}}$ $<$ 1119~\mevcsq\ while we reconstruct the $\Lambda_c^+$ decays involving $\Lambda$ directly in the final state and $\Sigma^{0}$ decays. 
%%\newline

\begin{figure}
\begin{center}
\includegraphics[width=3.5in]{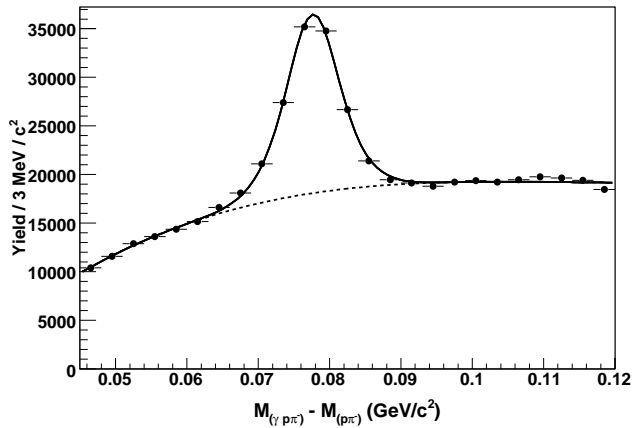}
\caption{The distribution of the invariant mass difference between $\gamma p\pi^{-}$ combinations and $p\pi^{-}$ candidates. The solid line indicates the result of the fit for the sum of the signal and background and the dashed line for the background only.}
\label{fig:sigma}
\end{center}
\end{figure}

The $\Sigma^{0}$ candidates are reconstructed in the decay mode $\Sigma^{0} \to  \Lambda \gamma $ using the already selected $\Lambda$ sample and photons with an energy greater than 0.1~GeV. 
The mass difference ($M_{\gamma p\pi^{-}} - M_{p\pi^{-}}$) is shown in Fig.~\ref{fig:sigma}. The distribution is fitted with the sum of two Gaussian functions with a common mean for the signal contribution, with $\sigma_{1} = 2.00 \pm 0.18~\mevcsq$ (the width of the core Gaussian) and $\sigma_{2} = 5.01 \pm 0.38~\mevcsq$ (the width of the wider Gaussian). In this fit $f_{2}$ = 60\%, and a 3$^{\mathrm{rd}}$ order polynomial is used for the background. We obtain a resolution $\sigma_{\rm RMS} = 4.0~\mevcsq$ and a mass difference between the $\Sigma^{0}$ and $\Lambda$ of 77.64 $\pm$ 0.04 (stat.)~\mevcsq. We accept candidates with ($M_{\gamma p\pi^{-}} - M_{p\pi^{-}}$) within 10~\mevcsq~of the mean value. 
%mean (difference between the invariant masses) of 77.64 $\pm$ 0.04 (stat.)~\mevcsq. We accept candidates with ($M_\gamma {p\pi^{-}} - M_{p\pi^{-}}$) within 10~\mevcsq~of the mean value.%%\newline
%%\indent 

To suppress combinatorial and $B\bar{B}$ backgrounds, we introduce $x_{p}$ as a scaled momentum of a $\Lambda^+_c$ candidate, where $x_{p}$ = $p^{*}$/$p^{*}_{\rm max}$. Here $p^{*}$ is the reconstructed momentum of the $\Lambda^+_c$ and $p^{*}_{\rm max}$ = $\sqrt{s/4 - M^{2}}$ with $\sqrt{s}$ is the total CM energy and $M$ is the reconstructed mass of the $\Lambda^+_c$ candidate. Our search is limited to $x_{p} > $ 0.5 or $x_{p}  > $ 0.6, depending on the decay mode so as to avoid the combinatorial background that dominates at low $x_{p}$. 
%The cut was chosen carefully for each mode by looking at control sample of simulated events for these decays modes.
%Charmed baryons at \pep2\ are produced either from the secondary decays of $B$ mesons or from \epem\ annihilaions to \ccbar\ jets.
%%%%%%%%%%%%%%%%%%%%%%%%%%%%%%%%%%%%%%%%%%%%%%%%
%Charmed baryons from $B$ meson decays are kinemetically limited to $x_{p} < $ 0.4, so our sear%ch is limited to  $\Lambda^+_c$ baryons produced in the \epem\ continuum.

%\indent To suppress combinatorial and $B\bar{B}$ backgrounds, we require
% $\Lambda^+_c$ candidates to have scaled momenta $x_{p}$ = $p^{*}$/$p^{*}_{\rm max} > $ 0.5. Here $p^{*}$ is the reconstructed momentum of the $\Lambda^+_c$ 
%candidate in the \epem CM frame, $p^{*}_{\rm max}$ = $\sqrt{s/4 - M^{2}}$, $\sqrt{s}$ is the total CM energy and $M$ is the reconstructed mass of the $\Lambda^+_c$ candidate. 

\section{PHYSICS RESULTS}
\label{sec:Physics}

\subsection{Study of the decays $\Lambda^+_c \to \Lambda K^+$ and $\Lambda^+_c \to \Sigma^{0}K^+$}
\label{sec:Measurement of LS}
\begin{figure}[!htb]
\begin{center}
\includegraphics[width=3.5in]{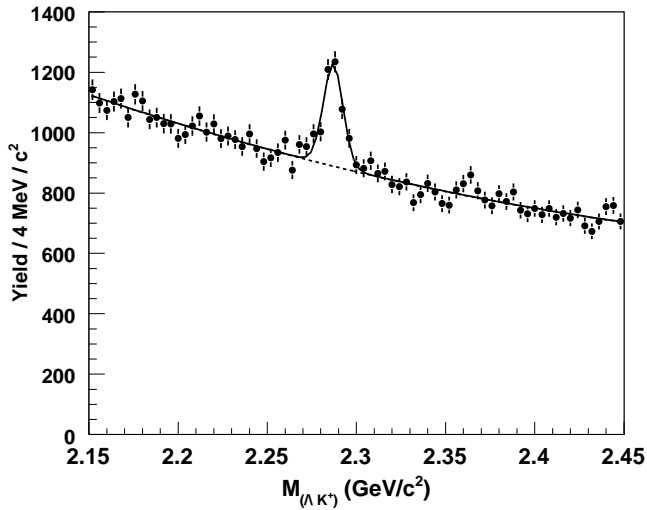}
\caption{The invariant mass of the $\Lambda K^+$ combinations for $x_{p}~>$ 0.5. The solid line indicates the result of the fit for the sum of the signal and background and the dashed line for the background only.}
\label{fig:lctolk}
\end{center}
\end{figure}

The reconstructed $\Lambda$ candidates are combined with a \Kp\ with the requirement $x_{p} > $ 0.5 to produce the mass spectrum shown in Fig.~\ref{fig:lctolk}. A clear \lcpl\ signal can be seen.
%to produce the mass spectrum and require $x_{p} > $ 0.5. A clear \lcpl\ signal can be seen in Fig.~\ref{fig:lctolk}. 
The mass distribution is fitted with a Gaussian function for the signal, and a 2$^{\mathrm{nd}}$ order polynomial for combinatorial background. The fit has a $\chi^2$ of 71.7 for 69 degrees of freedom. We obtain a raw yield of 1162 $\pm$ 101 (\textnormal{stat.}) events and a fitted width $\sigma$  = 5.5 $\pm$ 0.7 (stat.)~\mevcsq, which is consistent with the resolution of 6.1 $\pm$ 0.1~\mevcsq\ determined from a sample of simulated $\Lambda^+_c \to \Lambda K^+$ signal events. The fitted mean value 2286.9 $\pm$ 0.6 \mevcsq\ is found to be in agreement with the measured $\Lambda^+_c$ mass 2286.46 $\pm$ 0.14 \mevcsq~\cite{PDG}.

\begin{figure}[!htb]
\begin{center}
\begin{tabular}{c}
\includegraphics[width=3.5in,height=2.5in]{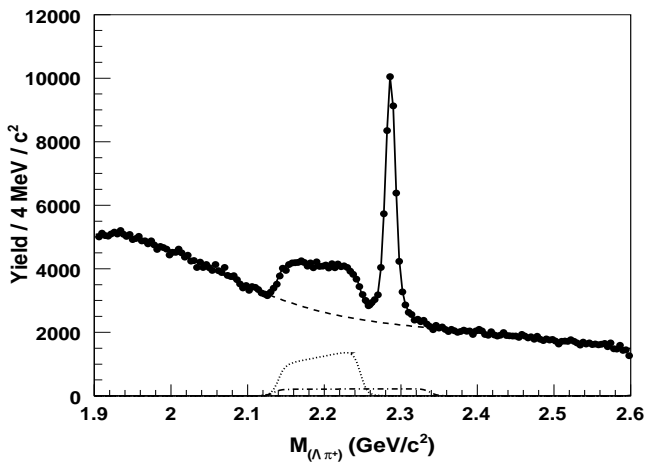}

\end{tabular}
\caption{The invariant mass of the $\Lambda \pi^+$ combinations for $x_{p}~>$ 0.5. The solid line indicates the result of the fit for the sum of the signal and backgrounds and the dashed line for the combinatorial background. The dotted line indicates the broad region corresponds to $\Lambda^+_c \to \Sigma^{0}\pi^+$ with a missing $\gamma$ and the dashed-dotted line represents the region corresponding to $\Xi_c^{+}/\Xi_c^{0}$ reflection with missing $\pi^{0}/\pi^{-}$.}
\label{fig:lctolp}
\end{center}
\end{figure}

\begin{figure}[!htb]
\begin{center}
\includegraphics[width=3.5in,height=2.5in]{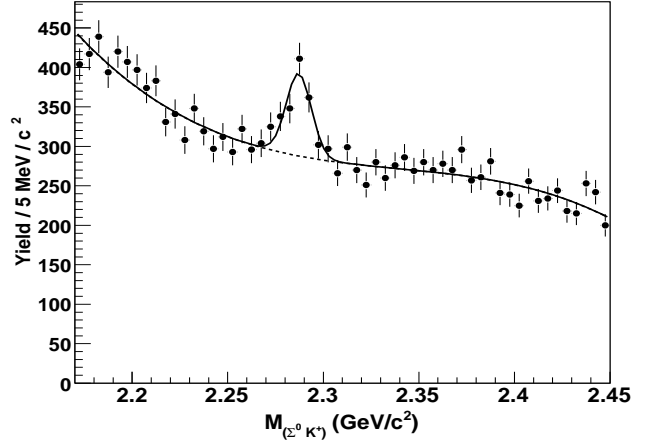}
\caption{The invariant mass of $\Sigma^{0}(\Lambda\gamma)K^+$ combinations for $x_{p}~>$ 0.5, where we do not replace the $\Sigma^{0}$ mass with the PDG value of the $\Sigma^{0}$ mass, as compared to what we have done for our final fit in Fig.~\ref{fig:lctosk}. 
%same as Fig.~\ref{fig:lctoskNoPDG}, except that in the fit we allow the width to float. The Fit uses a Gaussian for the signal and a 3$^{\mathrm{rd}}$ order polynomial to represent the background. 
The solid line indicates the result of the fit for the sum of the signal and background and the dashed line for the background only.}
%%The fitted yield is  $322.7~\pm~64.2$ (stat.) events with measured width of $\sigma$ = 6.1 $\pm$ 1.5~\mevcsq. This fit has a $\chi^2$ of 51.8 for 49 degrees of freedom.}
\label{fig:lctoskNoPDGFloat}
\end{center}
\end{figure}

\begin{figure}[!htb]
\begin{center}
\includegraphics[width=3.5in,height=2.5in]{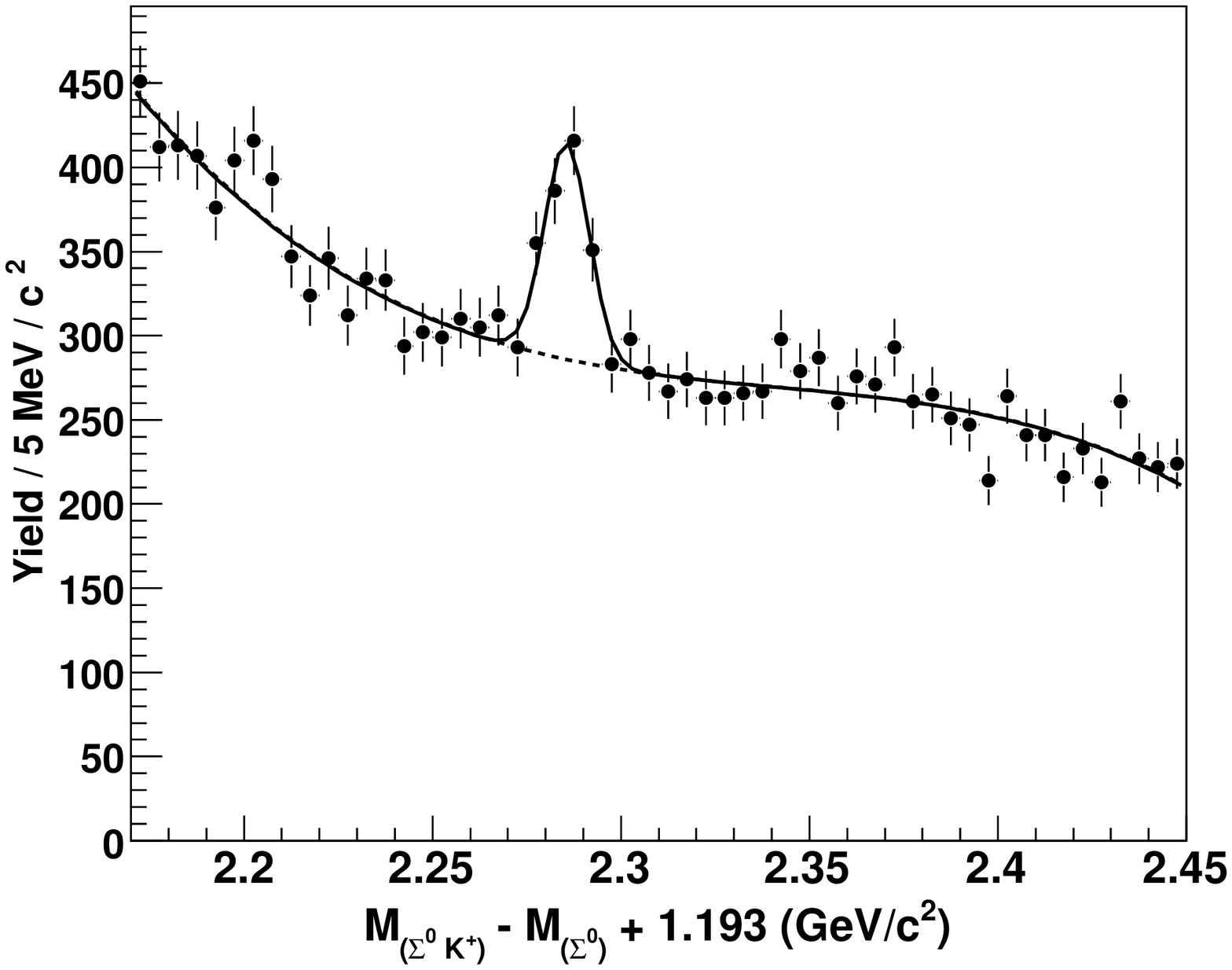}
\caption{The invariant mass of $\Sigma^{0}(\Lambda\gamma)K^+$ combinations for $x_{p}~>$ 0.5. The solid line indicates the result of the fit for the sum of the signal and background and the dashed line for the background only.}
\label{fig:lctosk}
\end{center}
\end{figure}

For reference, we use the decay $\Lambda^+_c \to \Lambda\pi^{+}$. The invariant mass distribution of $\Lambda \pi^{+}$ combinations is
shown in Fig.~\ref{fig:lctolp}. At mass values below the $\Lambda^+_c$ mass a broad distribution around 2.2 GeV/$c^{2}$ is visible. This peak corresponds to $\Lambda^+_c \to \Sigma^{0}\pi^+$ with a missing $\gamma$. Additionally, at 2.3 GeV/$c^{2}$ we see a shoulder, identified as the upper edge of a $\Xi_c^{+}/\Xi_c^{0}$ reflection with missing $\pi^{0}/\pi^{-}$. These shapes are established over the $\Lambda^+_c$ signal region using a large sample of \qqbar\ simulated events.
The distribution is fitted using the sum of two Gaussian functions with the same mean for the 
signal, a square wave function for each reflection, and a 7$^{\mathrm{th}}$ order polynomial for combinatorial background. Because of the presence of reflections (as described above) in this decay mode, we need to use a wider window (1.9 to 2.6 \gevcsq\ instead of 2.15 to 2.45 \gevcsq) to fit the background, and consequently a higher order polynomial to be able to reproduce the background shape over the extended range. The resultant $\chi^2$ of the fit to the data is 233.7 for 155 degrees of freedom. The fitted distribution gives a mean value of 2286.5 $\pm$ 0.1 \mevcc\, which is in agreement with the measured $\Lambda^+_c$ mass~\cite{PDG}. The fitted values for the width of the core Gaussian and, for the width of the wider Gaussian are $\sigma_1$ = 5.6 $\pm$ 0.1 \mevcc\ and, $\sigma_2$ = 11.6 $\pm$ 0.3 \mevcc\, respectively, with $f_{2} = 36\%$. We obtain a raw 
yield of 33543 $\pm$ 334 (stat.) events with the measured resolution $\sigma_{\rm RMS}$ = 8.3 $\pm$ 0.3 (stat.)~\mevcsq, which is consistent with the resolution of 8.0 $\pm$ 0.1 (stat.)$~\mevcsq$ measured from a sample of simulated $\Lambda^+_c \to \Lambda\pi^{+}$ signal events. 
Using signal MC, the ratio of signal reconstruction efficiencies $\epsilon$ is found to be ${\epsilon ( \Lambda^+_c \to \Lambda K^{+})}/{\epsilon ( \Lambda^+_c \to \Lambda\pi^{+})}$ = 0.781~$\pm$~0.010 (stat.). With this value we calculate:
\[\frac{\brc(\lcpl \ra \Lambda \kpl)}{\brc(\lcpl \ra \Lambda \pipl )} =~0.044~\pm~0.004~(\textnormal{stat.})~\pm~0.003~(\textnormal{syst.})~.\]
As a cross-check, we calculate the ratio {\brc(\lcpl \ra $\Lambda$ \kpl)}/{\brc(\lcpl \ra $\Lambda$ \pipl )} in on-resonance and off-resonance data separately. The value obtained using on- and off-peak samples agree within uncertainties (the ratio of on-peak to off-peak branching ratio is: $1.04~\pm~0.04$). 
We provide a detailed description of the sources of systematic uncertainty 
in Sec.~\ref{sec:Systematics}. 
\newline
\indent We also use the invariant mass distribution of $\Lambda \pi^{+}$ as shown in Fig.~\ref{fig:lctolp} to extract the yield of $\Lambda^+_c \to \Sigma^{0}\pi^+$ assuming a missing $\gamma$. The signal yield which is extracted from the corresponding square wave function fit is found to be: $32693 \pm 324$. We generate signal MC samples of $\Lambda^+_c \to \Sigma^{0}{\pipl}$ with missing $\gamma$ to evaluate the signal detection efficiency and to get the signal shape. The systematic uncertainty due to this is considered and is included as a part of total systematic uncertainty for this branching ratio. The relative  signal reconstruction efficiency from the MC is found to be ${\epsilon ( \Lambda^+_c \to \Sigma^{0}{\pipl})}/{\epsilon ( \Lambda^+_c \to \Lambda\pi^{+})}$ = 1.013~ $\pm$ ~0.010 (stat.). We measure \[\frac{\brc(\lcpl \ra \Sigma^{0} \pipl)}{\brc(\lcpl \ra \Lambda \pipl )} = ~0.977~ \pm ~0.015 ~(\textnormal{stat.})~ \pm ~0.051 ~(\textnormal{syst.})~.\]
\indent  
We combine the reconstructed \sigz\ candidates with a \Kp\ to form $\Lambda^+_c$ candidates and require $x_{p}>$ 0.5. We improve the invariant mass resolution by about 20$~\%$ by using the variable, $M_{\Sigma^{0}K^{+}} - M_{\Sigma^0} + M^{PDG}_{\Sigma^0}$, instead of $M_{\Sigma^{0}K^{+}}$, where $M_{\Sigma^0}$ is the reconstructed mass of the ${\Sigma^0}$ and $M^{PDG}_{\Sigma^0}$ is the world average for the mass of the ${\Sigma^0}$~\cite{PDG}. This method is also used in other experiments to improve the mass resolution~\cite{Belle}. For demonstration purposes we also show, in Fig.~\ref{fig:lctoskNoPDGFloat}, the $\Lambda^+_c \to \Sigma^{0}K^{+}$ mass distribution, where we do not replace the mass of $\Sigma^0$ by the fixed mass (PDG) value. The fit uses a Gaussian for the signal and a 3$^{\mathrm{rd}}$ order polynomial to represent the background. The fitted yield is  $323~\pm~64$ (stat.) events with measured width of $\sigma$ = 6.1 $\pm$ 1.5~\mevcsq. This fit has a $\chi^2$ of 51.8 for 49 degrees of freedom. The final fit for the invariant mass distribution of \sigz \Kp\ combinations is shown in Fig.~\ref{fig:lctosk}. 
%A clear \lcpl\ signal can be seen. 
 An attempt to fit the $\Lambda^+_c$ mass distribution to the sum of a single Gaussian and a 2$^{\mathrm{nd}}$ order polynomial shape yields a high $\chi^2$ of 72.4 for 50 degrees of freedom. However, if the fit is performed using a single Gaussian function 
%with a fixed width $\sigma$ = 6.0~\mevcsq\ (as determined from a sample of simulated $\Lambda^+_c \to \Sigma^{0}K^{+}$ signal events
%~\footnote{For demonstration purposes, we also show the distributions if the width is floated, for the mass difference method, Fig.~\ref{fig:lctoskPDGFloat}, and also when we do not replace the mass of $\Sigma^{0}$ with its PDG value~\ref{fig:lctoskNoPDGFloat}}) 
for the signal and a 3$^{\mathrm{rd}}$ order polynomial for combinatorial background, the resultant $\chi^2$ is 47.8 for 49 degrees of freedom. The fit yields 366 $\pm$ 52 (stat.) events. The measured width $\sigma$ = 5.7 $\pm$ 0.8~\mevcsq\ is consistent with the resolution $\sigma$ = 6.0 $\pm$ 0.1~\mevcsq\ determined from a sample of simulated $\Lambda^+_c \to \Sigma^{0} K^{+}$ signal events. The fitted mean value 2286.0 $\pm$ 0.9 \mevcsq\ is in agreement with the measured $\Lambda^+_c$ mass~\cite{PDG}. 
%%%%The fit yields 376 $\pm$ 45 (stat.) events. 

\begin{figure}[!htb]
\begin{center}
\begin{tabular}{c}
\mbox{\includegraphics[width=3.5in]{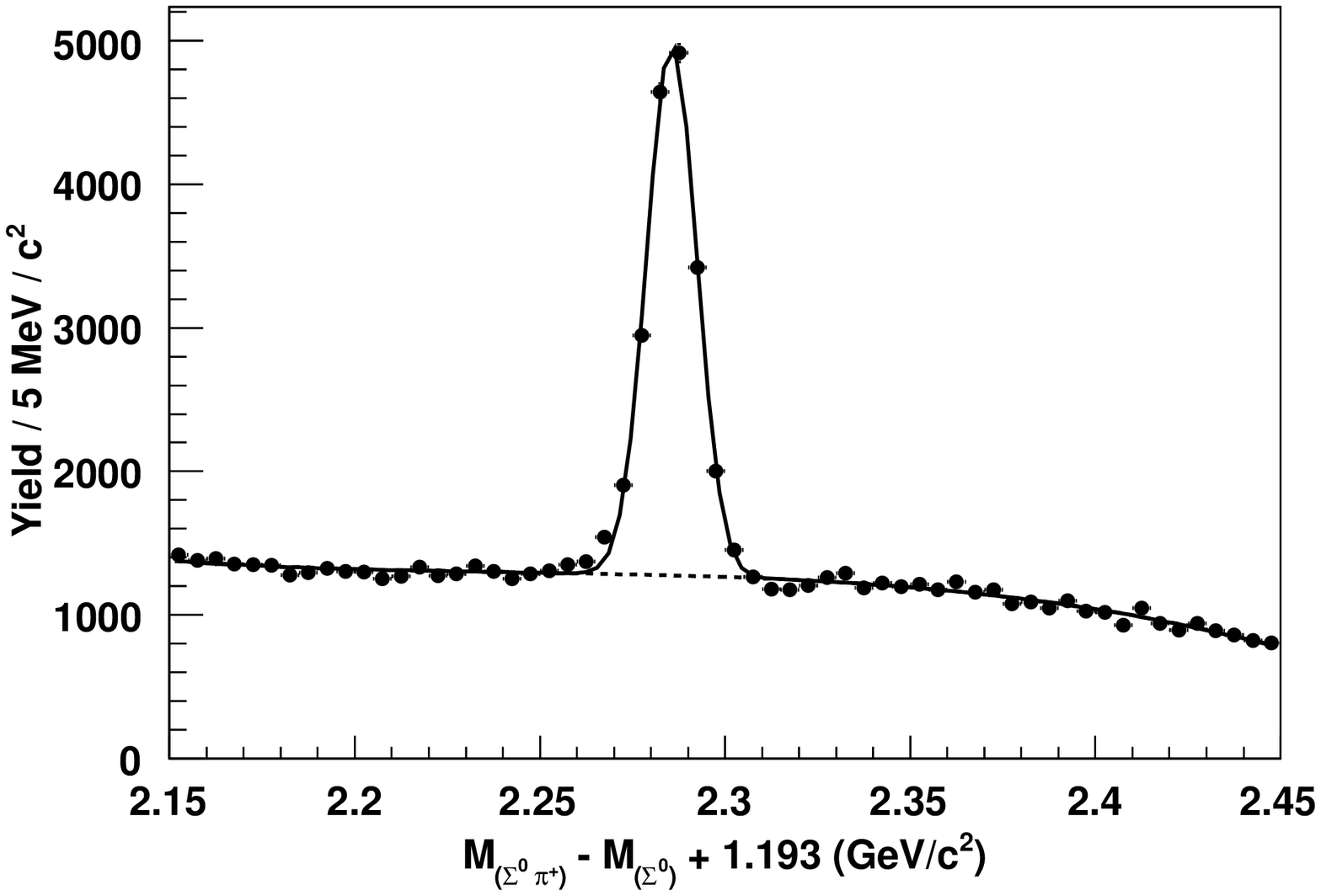}}
\end{tabular}
\caption{The invariant mass of $\Sigma^{0}(\Lambda\gamma)\pi^+$ combinations for $x_{p}~>$ 0.5. The solid line indicates the result of the fit for the sum of the signal and background and the dashed line for the background only.}
\label{fig:lctosp}
\end{center}
\end{figure}

For reference, we use the Cabibbo-favored decay mode $\Lambda^+_c \to \Sigma^{0}\pi^{+}$. The invariant mass of the $\Sigma^{0}\pi^{+}$ combinations is shown in Fig.~\ref{fig:lctosp}. An attempt to fit this distribution to a sum of single Gaussian and a 2$^{\mathrm{nd}}$ order polynomial gives a $\chi^2$ of 119.9 for 54 degrees of freedom, which is not the best choice for this fit. However, the final fit uses a Gaussian function for the signal and a 3$^{\mathrm{rd}}$ order polynomial for background, gives a $\chi^2$ of 87.3 for 53 degrees of freedom. The fit yields $12490~\pm~162$ (stat.) events. The measured width of $\sigma$ = 6.7 $\pm$ 0.1~\mevcsq\ is consistent with the resolution $\sigma$ = 7.1 $\pm$ 0.1~\mevcsq\ measured in a sample of simulated $\Lambda^+_c \to \Sigma^{0}\pi^{+}$ signal events. The fitted mean value 2285.6 $\pm$ 0.7 \mevcsq\ is also in agreement with the measured $\Lambda^+_c$ mass~\cite{PDG}. The relative reconstruction efficiency
is measured to be ${\epsilon ( \Lambda^+_c \to \Sigma^{0}K^{+})}/{\epsilon ( \Lambda^+_c \to \Sigma^{0}\pi^{+})}$ = 0.780 ~$\pm$ ~0.010 (stat.) using signal MC samples. The resulting relative branching ratio is 
\[\frac{\brc(\lcpl \ra \Sigma^{0} \kpl)}{\brc(\lcpl \ra \Sigma^{0} \pipl )} =~0.038~\pm~0.005~(\textnormal{stat.})~\pm~0.003~(\textnormal{syst.})~ .\]


\begin{figure}[!htb]
\begin{center}
\begin{tabular}{c}
\mbox{\includegraphics[width=3.5in]{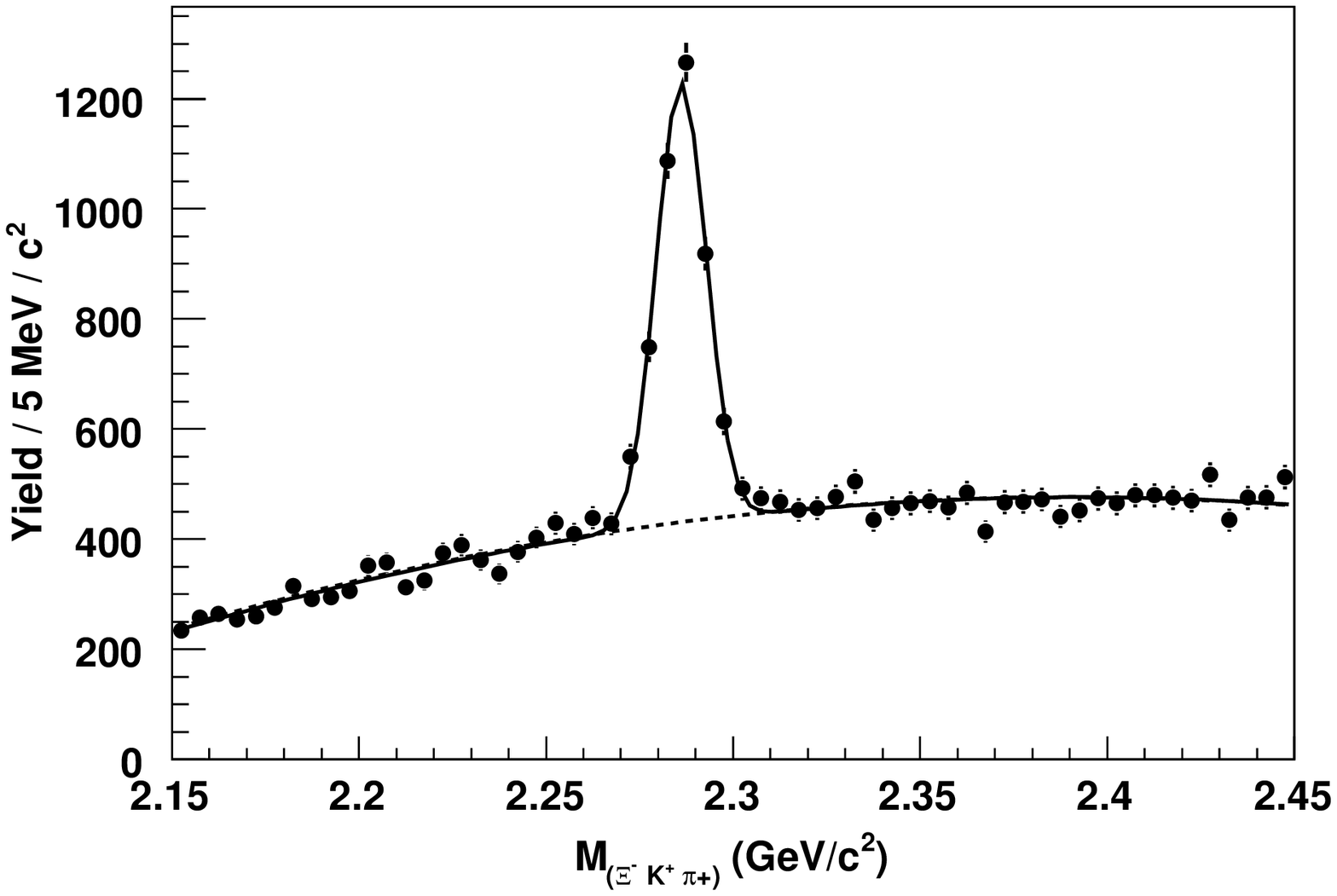}}
\end{tabular}
\caption{ Combinatorial ${\casmi}(\Lambda \pi^{-})K^+\pi^{+}$ invariant mass distribution for $x_{p}~>$ 0.6. The solid line indicates the result of the fit for the sum of the signal and background and the dashed line for the background only.}
\label{fig:lctocaskp}
\end{center}
\end{figure}
\begin{figure}[!htb]
\begin{center}
\begin{tabular}{c}
\mbox{\includegraphics[width=3.5in]{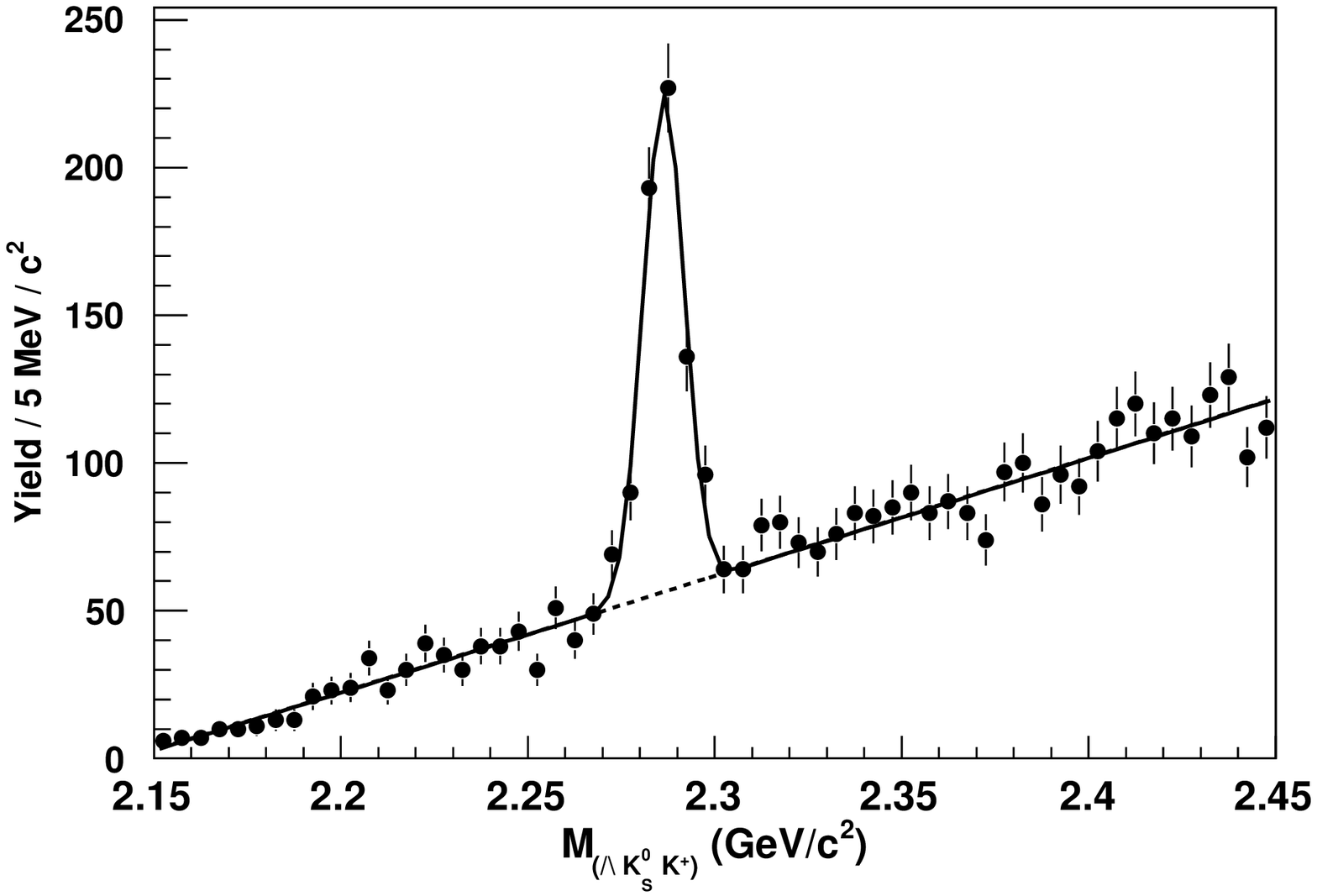}}
\end{tabular}
\caption{ Combinatorial $\Lambda {\kzsh}(\pi^{+}\pi^{-}) K^+$ invariant mass distribution for $x_{p}~>$ 0.6. The solid line indicates the result of the fit for the sum of the signal and background and the dashed line for the background only.}
\label{fig:lctolksk}
\end{center}
\end{figure}

\begin{figure}[!htb]
\begin{center}
\begin{tabular}{c}
\mbox{\includegraphics[width=3.5in]{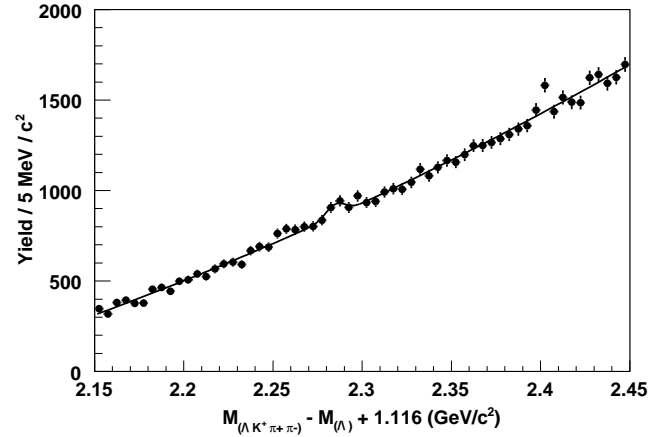}}
\end{tabular}
\caption{The invariant mass of  $\Lambda K^+\pi^{+}\pi^{-}$ combinations for $x_{p}~>$ 0.6. The solid line indicates the result of the fit for the sum of the signal and background.} 
\label{fig:lctolkpp}
\end{center}
\end{figure}
\subsection{Search for the decay of $\Lambda^+_c \to \Lambda K^+\pi^{+}\pi^{-}$ }
%\indent To measure the Cabibbo-suppressed decay $\Lambda^+_c \to\Lambda K^{+}\pi^{+}\pi^{-}$ we use the selection criteria described in Sec.~\ref{sec:Analysis}, with the scaled momentum restricted to $x_{p} >$ 0.6 in order to further reduce the combinatorial background. For the reference we use the $\Lambda^+_c \to \Lambda \pi^{+}$ decay mode with the same scaled momentum selection, for which we obtain a raw yield of 22204 $\pm$ 257 (stat.) events. 
\indent To measure the Cabibbo-suppressed decay $\Lambda^+_c \to\Lambda K^{+}\pi^{+}\pi^{-}$ we use the selection criteria described in Sec.~\ref{sec:Analysis}. This decay mode has multiple particles in the final state. The combinatorial background is relatively higher here than in the processes like $\Lambda^+_c \to \Lambda \pi^{+}/K^+$. The scaled momentum is restricted to $x_{p} >$ 0.6 in order to reduce the combinatorial background. For the reference we use the $\Lambda^+_c \to \Lambda \pi^{+}$ decay mode with the same scaled momentum selection, for which we obtain a raw yield of 22204 $\pm$ 257 (stat.) events. 
%\newline
%\indent 

We search for all the possible decays leading to the 
same final state as $\Lambda K^+\pi^{+}\pi^{-}$ and find that the major contributions come from the decays of $\Lambda^+_c \to {\casmi}K^+\pi^{+}$ (${\casmi} \to \Lambda \pi^{-}$)  and $\Lambda^+_c \to \Lambda {\kzsh} K^+$ (${\kzsh} \to \pi^{+}\pi^{-}$). We confirmed these contributions using the MC truth matching for our continuum MC. We reconstruct a {\casmi} candidate from a $\Lambda$ candidate and a $\pi^{-}$ track requiring an invariant mass within 15~\mevcsq\ around the nominal value 1321.3~\mevcsq\ \cite{PDG}. The invariant mass distribution of ${\casmi}K^+\pi^{+}$ combinations is shown in Fig.~\ref{fig:lctocaskp}. The distribution is fitted with a single Gaussian for the signal and a 2$^{\mathrm{nd}}$ order polynomial for the background, with a resultant $\chi^2$ of 67.5 for 54 degrees of freedom. We obtain a width $\sigma = 6.6 \pm 0.2~\mevcsq$  and a signal yield of 2665 $\pm$ 84 (stat.). The relative signal reconstruction efficiency is measured to be ${\epsilon ( \Lambda^+_c \to {\casmi}K^{+}\pi^{+})}/{\epsilon ( \Lambda^+_c \to \Lambda \pi^{+})}$ = 0.250 ~$\pm$~ 0.003 (stat.). Accounting for the $\casmi$ sub-decay branching fraction~\cite{PDG}, the branching ratio is measured to be \[\frac{\small{\brc(\lcpl\to\casmi\kpl\pipl} )}{\small{\brc(\lcpl \ra \Lambda \pipl)}}= 0.480 \pm 0.016 (\textnormal{stat.}) \pm 0.038 (\textnormal{syst.}).\]

\indent We also reconstruct {\kzsh} candidates formed from two tracks identified as a {\pipl} and a {\pimi} with invariant mass 489 $<$ $M_{\pi\pi}$ $<$ 509 ~{\mevcsq}. The invariant mass distribution of $\Lambda^+_c \to {\Lambda} {\kzsh} K^+$ is shown in Fig.~\ref{fig:lctolksk}. The fit is performed using a single Gaussian for the $\Lambda^+_c$ signal whereas the background is described by a 2$^{\mathrm{nd}}$ order polynomial function. The resultant fit has a $\chi^2$ of 43.5 for 54 degrees of freedom. The measured width is 5.5 $\pm$ 0.4 (stat.)~{\mevcsq} and a signal yield of 460 $\pm$ 30 (stat.) is obtained. Using a signal reconstruction efficiency of ${\epsilon ( \Lambda^+_c \to {\Lambda} {\kzsh}K^{+})}/{\epsilon ( \Lambda^+_c \to \Lambda \pi^{+})}$ = 0.152 ~$\pm$~ 0.020 (stat.) and accounting for the $\kzb$ ($\kzb \to \kzsh$) and $\kzsh$ ($\kzsh \to \pi^{+}\pi^{-}$) sub-decay branching fractions ~\cite{PDG}, the branching ratio is measured to be  \[\frac{\small{\brc(\lcpl \to \Lambda \kzb \kpl )}}{\small{\brc(\lcpl \ra \Lambda \pipl )}} = 0.395 \pm 0.026 (\textnormal{stat.}) \pm 0.036 (\textnormal{syst.}).\] 

\indent We reject the contribution from the above Cabibbo-favored decay modes by excluding the ${\casmi}$ and ${\kzsh}$ mass windows as mentioned above. The final invariant mass distribution of $\Lambda K^+\pi^{+}\pi^{-}$ combinations is shown in Fig.~\ref{fig:lctolkpp}. We fit the mass distribution using a Gaussian function for the signal and a 2$^{\mathrm{nd}}$ order polynomial for the combinatorial background. We fix the width $\sigma$ = 5.2 \mevcsq\ and the mean 2285.5 \mevcsq\ as predicted from a sample of simulated signal events for this decay. We obtain a signal yield of 158 $\pm$ 63 (stat.) events for the $\Lambda^+_c \to \Lambda K^+\pi^{+}\pi^{-}$ decay. The goodness for this fit shows a $\chi^2$ of 56.6 for 56 degrees of freedom. The relative signal reconstruction efficiency is measured to be ${\epsilon ( \Lambda^+_c \to \Lambda K^{+}\pi^{+}\pi^{-})}/{\epsilon ( \Lambda^+_c \to \Lambda \pi^{+})}$ = 0.310 ~$\pm$~ 0.010 (stat.). Since we do not observe a statistically significant signal for $\Lambda^+_c \to \Lambda K^{+}\pi^{+}\pi^{-}$, we calculate the upper limit at $90~\%$ confidence level (C.L.) 
 using the Feldman and Cousins method~\cite{feld} and including systematic uncertainties. \[\frac{\brc(\lcpl \ra \Lambda \kpl \pipl \pimi)}{\brc(\lcpl \ra \Lambda \pipl )} < 4.1 \times ~10^{-2}  ~@ ~90~\% ~{\rm {C.L.}}\] 
\subsection{Search for the decay of $\Lambda^+_c \to \Sigma^{0}K^+\pi^{+}\pi^{-}$}
\begin{figure}
\begin{center}
\begin{tabular}{c}
\mbox{\includegraphics[width=3.5in]{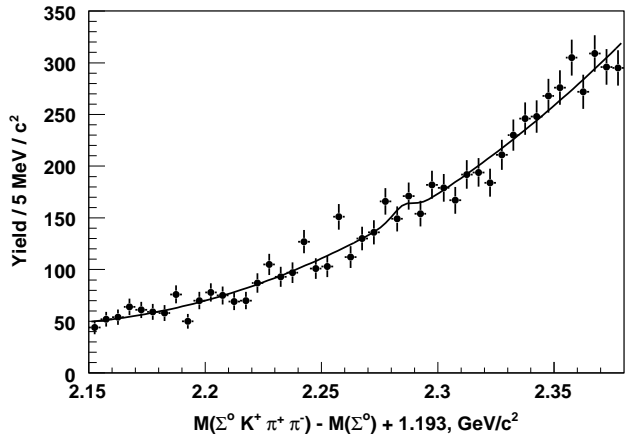}}
\end{tabular}
\caption{The invariant mass of $\Sigma^{0}(\Lambda\gamma)K^+\pi^{+}\pi^{-}$ combinations for $x_{p}~>$ 0.6. The solid line indicates the result of the fit for the sum of the signal and background.}
\label{fig:lctoskpp}
\end{center}
\end{figure}

\indent We search for the decay $\Lambda^+_c \to \Sigma^{0}K^+\pi^{+}\pi^{-}$ using the selection described in Sec.~\ref{sec:Analysis} and restricting the scaled momentum to $x_{p} >$ 0.6. The invariant mass distribution of $\Sigma^{0}K^+\pi^{+}\pi^{-}$ is shown in Fig.~\ref{fig:lctoskpp}. The $\Lambda^+_c$ mass distribution is fitted using a single Gaussian function with fixed width $\sigma$ = 4.4~\mevcsq\ and mean = 2284.7~\mevcsq\ (measured from a sample of simulated signal events for this decay) for the signal and a 2$^{\mathrm{nd}}$ order polynomial for combinatorial background, with a $\chi^2$ of 48.9 for 41 degrees of freedom. The fit yields 21 $\pm$ 24 (stat.) events. Using the decay mode $\Lambda^+_c \to \Sigma^{0}\pi^{+}$ for reference, we find a raw yield of 8848 $\pm$ 126 (stat.) events for this decay in the range $x_{p} >$ 0.6. The relative reconstruction efficiency is determined to be ${\epsilon ( \Lambda^+_c \to \Sigma^{0}K^{+}\pi^{+}\pi^{-})}/{\epsilon ( \Lambda^+_c \to \Sigma^{0}\pi^{+})}$ = 0.390 ~$\pm$~0.010 (stat.). 
We do not observe a statistically significant signal for $\Lambda^+_c \to \Sigma^{0}K^+\pi^{+}\pi^{-}$. We calculate the upper limit using the Feldman and Cousins method~\cite{feld} and including systematic uncertainties. We find:
\[\frac{\brc(\lcpl \ra \Sigma^{0} \kpl \pipl \pimi)}{\brc(\lcpl \ra \Sigma^{0} \pipl )} < 2.0 \times ~10^{-2}  ~@ ~90~\% ~{\rm {C.L.}}\]

\section{SYSTEMATIC STUDIES}
\label{sec:Systematics}
\indent We consider several possible sources of systematic uncertainties in our measurements, as shown in Table~\ref{tab:Syst1}. The systematic uncertainty due to limited signal MC statistics is between 1$\%$ and 3$\%$ depending on the decay mode. The systematic uncertainty due to each requirement in the candidate selection is estimated by varying the selection cuts (mass window for the resonance masses, cuts on $\Lambda$ flight distance and vertex $\chi^{2}$, $x_{p}$ cut). It is typically about $1\%$ and always below $4\%$. For ${\brc(\Lambda^+_c \to \Sigma^{0}K^{+})} /{\brc(\Lambda^+_c \to \Sigma^{0}\pi^{+})}$, the photon spectrum is different in the signal and reference decay modes, leading to a systematic uncertainty of less than 1$\%$ obtained by changing the photon energy cut in both modes. The uncertainty due to track finding is about 2.8 $\%$ for modes with higher multiplicity than the reference decay mode. The systematic uncertainty due to a $\pi^{\pm}$ misidentified as a $K^{\pm}$ is below 1$\%$. By studying large signal MC samples the change in detection efficiency with and without a vertex requirement for $\kzsh \to \pi^{+} \pi^{-}$ and $\casmi \to \Lambda \pi^{-}$, we assign a systematic uncertainty due to the lack of vertexing for {\kzsh} and {\casmi} to be $5\%$ for the modes with such a particle in the final state. We use a simplistic phase space model to generate signal MC for $\Lambda^+_c \to \Lambda K^+\pi^{+}\pi^{-}$ and $\Lambda^+_c \to \Sigma^{0}K^+\pi^{+}\pi^{-}$. We assign 5.4 $\%$ for signal MC modeling.
 
We also study possible biases due to our fitting procedure by varying the fitted function as describled below. Conservatively, the effect of all changes to the signal yield are accounted as systematic uncertainty. For each distribution we vary the order of the polynomial background, and vary the signal width ($\sigma$) by one standard deviation. In addition, for $\Lambda^+_c \to \Lambda \pi^{+}$ we vary all the parameters describing the $\Sigma^{0}$ and $\Xi_c$ reflections by one standard deviation. The systematic uncertainty due to fit bias is assigned to be 5.9~$\%$ for ${\brc(\Lambda^+_c \to \Lambda K^{+})}/{\brc(\Lambda^+_c \to \Lambda \pi^{+})}$, 8$~\%$ for ${\brc(\Lambda^+_c \to \Sigma^{0}K^{+})}/{\brc(\Lambda^+_c \to \Sigma^{0}\pi^{+})}$, and 4.7$~\%$ for ${\brc(\Lambda^+_c \to \Lambda K^{+}\pi^{+}\pi^{-})}/{\brc(\Lambda^+_c \to \Lambda \pi^{+})}$. The systematic uncertainty associated with the fitting is found to be the dominant one for the Cabibbo-suppressed decay modes. Published data~\cite{PDG} provide the uncertainty in the daughter branching fractions ($1\%$ - $4\%$)

\begin{table*}[h!]

\caption{Summary of sources of systematic uncertainties ($\%$).}

\begin{center}
\tabcolsep=3mm
\begin{tabular}{ l c c c c c c c}\hline\hline

 Sources of uncertainty  & $\Lambda$ \kpl\ & $\Lambda$ \kpl \pipl \pimi & \sigz \kpl\  & \sigz \kpl \pipl \pimi &\sigz \pipl\ &  \casmi \kpl \pipl&$\Lambda$ \kzb \kpl\ \\ 
\hline
MC statistics            & 1.1  & 1.9 & 1.6 & 2.4 & -&2.0&2.1\\ 
\hline
$\Lambda_{mass}$          & 0.6 & 0.1 & 1.2 & 1.2 &-&1.1&0.1\\ 
\hline
$P_{\chi^{2}}$             & 3.8 & 0.7 & 0.9 &0.9 &-& 0.7&0.7\\
\hline
$\Lambda$ flight          & 0.7 & 2.8 & 1.9 & 1.9 &-& 2.4&3.4\\
\hline
$x_{p}$                   & 0.7  & 1.8&2.1 & 2.1 &-& 2.2&1.8\\
\hline
$E_{\gamma}$               & - & - & 0.9 & 0.9 &-& -&-\\
\hline
Tracking                   & - & 2.8 &- & 2.8 &-& 2.8&2.8\\
\hline
Vertexing(\kzsh,\casmi)   & - & - &- & - &-& 5.0& 5.0\\
\hline
\sigz mass                & - & - & 1.3 & 1.3& -& -&-\\
\hline
\casmi mass                & - & 1.5 & -& -&- & 1.2& 2.6\\
\hline
\kzsh mass                 & - & 0.8 &- & - & -& 1.0&1.9\\
\hline
MC Modeling                & - & 5.4 & -& 5.4& -& -&-\\
\hline
Fitting                    & 5.9& 4.7 & 8.0 & 5.0 & 5.2&1.4&4.1\\
\hline                   
\BRcastlpi  & -& - & - &  & -&3.5&-\\
\hline
\BRkstpipi\  &  -& - & - &  & -&-&1.0\\
\hline 
Total Systematic           & 7.2& 8.8 & 8.9 &9.0& 5.2 & 8.1 &9.0\\

\hline\hline
\end{tabular}
\end{center}
\label{tab:Syst1}
\end{table*} 

\begin{table*}
\caption{Summary of signal yields, relative efficiencies and branching fraction ratios with respect to the reference mode for the Cabibbo-suppressed decays of $\Lambda_{c}^{+}$, where the first uncertainty is statistical and second one represents the systematic uncertainty. The decay $\lcpl \ra \Lambda \pipl$  is the reference mode for $\Lambda^+_c \to \Lambda K^{+}$ and $\Lambda^+_c \to \Lambda K^+\pi^{+}\pi^{-}$ signal decay modes. The decay $\lcpl \ra \Sigma^{0} \kpl$ is the reference mode for $\Lambda^+_c \to \Sigma^{0}K^{+}$ and $\Lambda^+_c \to \Sigma^{0}K^+\pi^{+}\pi^{-}$ signal decay modes.  }

\begin{center}
\tabcolsep=8mm
\begin{tabular}{ l  c  c  c}\hline\hline

  Signal Mode  & Signal Yield   & Relative  Efficiency  & ${\brc_{\rm signal}}$ / ${\brc_{\rm reference}}$ \\ 
\hline
 $\Lambda$ \kpl\   ($x_{p} > 0.5$)   & $1162 \pm 101$ & 0.781 $\pm$ 0.010 & $0.044 \pm 0.004 \pm 0.003$  \\ 
\hline

 $\Lambda$ \kpl \pipl \pimi  ($x_{p} > 0.6$)  & $160 \pm 62$  & 0.310 $\pm$ 0.010 & $< 4.1 \times 10^{-2}$  90$\%$ C.L. \\ 
\hline
\sigz \kpl\ ($x_{p} > 0.5$) & $366 \pm 52$ &  0.780 $\pm$ 0.010 & $0.038 \pm 0.005 \pm 0.003$   \\ 
%%%\sigz \kpl\ ($x_{p} > 0.5$) & $376 \pm 45$ &  0.780 $\pm$ 0.010 & $0.039 \pm 0.005 \pm 0.003$   \\ 
\hline
 \sigz \kpl \pipl \pimi ($x_{p} > 0.6$)  & $21 \pm 24$ & 0.390 $\pm$ 0.010 & $< 2.0 \times 10^{-2}$   90$\%$ C.L.    \\ 

\hline\hline
\end{tabular}
\end{center}
\label{tab:Summary1}
\end{table*} 

\begin{table*}[h!]

\caption{Summary of signal yields, relative efficiencies and branching fraction ratios with respect to the reference mode for the Cabibbo-favored decays of $\Lambda_{c}^{+}$, where the first uncertainty is statistical and the second one represents the systematic uncertainty. The decay $\lcpl \ra \Lambda \pipl$  is the reference mode for
 $\lcpl \ra \sigz \pipl$, $\lcpl \ra \casmi \kpl \pipl$ and $\lcpl \ra \Lambda \kzb \kpl$ signal decay modes.}

\begin{center}
\tabcolsep=8mm
\begin{tabular}{ l c c c   }\hline\hline

  Signal Mode  & Signal Yield & Relative Efficiency   & ${\brc_{\rm signal}}$ / ${\brc_{\rm reference}}$ \\ 
\hline
 \sigz \pipl\  ($x_{p} > 0.5$)    & $32693 \pm 324$ & 1.013 $\pm$ 0.010 & $0.977 \pm 0.015 \pm 0.051$  \\ 
\hline

  \casmi \kpl \pipl   ($x_{p} > 0.6$) & $2665 \pm 84$  & 0.250 $\pm$ 0.003 & $0.480 \pm 0.016 \pm 0.039$  \\ 
\hline

 $\Lambda$ \kzb \kpl\  ($x_{p} > 0.6$)  & $460 \pm 30$ &  0.152$\pm$ 0.020 & $0.395 \pm 0.026 \pm 0.036$  \\

\hline\hline
\end{tabular}
\end{center}
\label{tab:Summary2}
\end{table*} 
\section{SUMMARY}
\label{sec:Summary}
We measure the branching ratio of the Cabibbo-suppressed decay  $ \Lambda^+_c \to \Lambda K^{+}$ relative to the Cabibbo-favored decay mode $\Lambda^+_c$ $\to$ $\Lambda \pi^+$ to be $ 0.044 \pm 0.004 (\textnormal{stat.}) \pm 0.003 (\textnormal{syst.})$, which is somewhat lower and substantially more precise than the previous measurement, 0.074 $\pm$ 0.010(stat.) $\pm$ 0.012(syst.)~\cite{Belle}. We also report the branching ratio of the Cabibbo-suppressed decay $\Lambda^+_c \to \Sigma^{0}K^{+}$ relative to the Cabibbo-favored decay mode $\Lambda^+_c$ $\to$ $\Sigma^{0} \pi^+$ to be $ 0.038 \pm 0.005 (\textnormal{stat.}) \pm 0.003 (\textnormal{syst.})$. It is also lower and substantially more precise than the previous measurement, 0.056 $\pm$ 0.014 $\pm$ 0.003~\cite{Belle}. 
We also report the first searches for the Cabibbo-suppressed decays $\Lambda^+_c \to \Lambda K^+\pi^{+}\pi^{-}$ and $\Lambda^+_c \to \Sigma^{0}K^+\pi^{+}\pi^{-}$. We do not observe statistically significant signals for these decay modes and therefore set upper limits at the 90$~\%$~{\rm {C.L.}} The results for the Cabibbo-suppressed decays are shown in Table~\ref{tab:Summary1}. We finally report the branching ratio measurement of the Cabibbo-favored decays $\Lambda^+_c$ $\to$ $\Sigma^{0} \pi^+$, $ \Lambda^+_c \to {\casmi} K^{+}{\pipl}$ and $\Lambda^+_c \to \Lambda \kzb K^{+}$ relative to the Cabibbo-favored decay mode $\Lambda^+_c$ $\to$ $\Lambda \pi^+$ as shown in Table~\ref{tab:Summary2}. These results represent a marked improvement on the existing numbers~\cite{PDG} and the results for the two body decays are also in agreement with the predictions for these modes. The expectations from the quark model~\cite{Khana} are ${\brc(\lcpl \ra \Lambda \kpl)}/{\brc(\lcpl \ra \Lambda \pipl )} = [0.039-0.056]$ and ${\brc(\lcpl \ra \Sigma^{0} \kpl)}/{\brc(\lcpl \ra \Sigma^{0} \pipl )} = [0.033-0.036]$.
%\begin{figure}[!htb]
%\begin{center}
%\includegraphics[width=3.5in,height=2.5in]{PRD_PLOT/lctosigmakPDGFloat.eps}
%\caption{The invariant mass of $\Sigma^{0}(\Lambda\gamma)K^+$ combinations for $x_{p}~>$ 0.5. The data are the same as in Fig.~\ref{fig:lctosk}, but in the fit we allow the width to float. The Fit uses a Gaussian for the signal and a 3$^{\mathrm{rd}}$ order polynomial to represent the background. The solid line indicates the result of the fit for the sum of the signal and background and the dashed line for the background only. The fitted yield is  $365.5~\pm~52.4$ (stat.) events and a measured width of $\sigma$ = 5.7 $\pm$ 0.8~\mevcsq. The fit has a $\chi^2$ of 47.8 for 49 degrees of freedom.}
%\label{fig:lctoskPDGFloat}
%\end{center}
%\end{figure}
%\begin{figure}[!htb]
%\begin{center}
%\includegraphics[width=3.5in,height=2.5in]{PRD_PLOT/lctosigmakNoPDG.eps}
%\caption{The invariant mass of $\Sigma^{0}(\Lambda\gamma)K^+$ combinations for $x_{p}~>$ 0.5, where we do not replace the $\Sigma^{0}$ with the PDG value of the $\Sigma^{0}$ mass. The Fit uses a Gaussian for the signal and a 3$^{\mathrm{rd}}$ order polynomial to represent the background. The solid line indicates the result of the fit for the sum of the signal and background and the dashed line for the background only. The fitted yield is $318.6~\pm~43.8$ (stat.) events and the fixed width $\sigma$ = 6.0~\mevcsq\ (as determined from a fit to simulated signal events). This fit has a $\chi^2$ of 51.9 for 50 degrees of freedom.}
%\label{fig:lctoskNoPDG}
%\end{center}
%\end{figure}
%\vspace{-0.25in}
\begin{acknowledgements}
\input pubboard/acknowledgements.tex
\end{acknowledgements}

%\pagebreak

\end{document}

%% file: pubboard/authors_sep2005.tex
%% author list as of 02-Sep-2005 (633 authors)
%
\author{B.~Aubert}
\author{R.~Barate}
\author{D.~Boutigny}
\author{F.~Couderc}
\author{Y.~Karyotakis}
\author{J.~P.~Lees}
\author{V.~Poireau}
\author{V.~Tisserand}
\author{A.~Zghiche}
\affiliation{Laboratoire de Physique des Particules, F-74941 Annecy-le-Vieux, France }
\author{E.~Grauges}
\affiliation{IFAE, Universitat Autonoma de Barcelona, E-08193 Bellaterra, Barcelona, Spain }
% ...
\collaboration{The \babar\ Collaboration}
\noaffiliation

%% file: abstract.tex
\indent We measure the branching ratios of the Cabibbo-suppressed decays $\Lambda^+_c$ $\to$  $\Lambda$ $K^+$ and $\Lambda^+_c$ $\to$ $\Sigma^{0}$ $K^+$
%(measured with improved accuracy). 
relative to the Cabibbo-favored decay modes $\Lambda^+_c$ $\to$ $\Lambda$ $\pi^+$ and $\Lambda^+_c$ 
$\to$ $\Sigma^{0}$ $\pi^+$ to be $ 0.044 \pm 0.004 ~(\textnormal{stat.})~ \pm ~0.003 ~(\textnormal{syst.})$ and $ 0.038~ \pm ~0.005 ~(\textnormal{stat.})~ \pm ~0.003 ~(\textnormal{syst.})$, respectively.
We set an upper limit on the branching ratio at the 90 $\%$ confidence level for $\Lambda^+_c$ $\to$ $\Lambda$ $K^+ \pi^+ \pi^-$ of $ 4.1 \times ~10^{-2}$ relative to $\Lambda^+_c$ $\to$ $\Lambda$ $\pi^+$, and for $\Lambda^+_c$ $\to$ $\Sigma^{0}$ $K^+ \pi^+ \pi^-$ of $ 2.0 \times ~10^{-2}$ relative to $\Lambda^+_c$ $\to$ $\Sigma^{0}$ $\pi^+$. We also measure the branching fraction for the Cabibbo-favored mode $\Lambda^+_c$ $\to$ $\Sigma^{0}$ $\pi^+$ relative to $\Lambda^+_c$ $\to$ $\Lambda$ $\pi^+$ to be $0.977~ \pm ~0.015 ~(\textnormal{stat.})~ \pm ~0.051 ~(\textnormal{syst.})$.  This analysis was performed using a data sample with an integrated luminosity of 125 fb$^{-1}$ collected by the $BABAR$ detector at the PEP-II asymmetric-energy $B$ factory at SLAC.

%% file: pubboard/acknowledgements.tex
We are grateful for the 
extraordinary contributions of our \pep2\ colleagues in
achieving the excellent luminosity and machine conditions
that have made this work possible.
The success of this project also relies critically on the 
expertise and dedication of the computing organizations that 
support \babar.
The collaborating institutions wish to thank 
SLAC for its support and the kind hospitality extended to them. 
This work is supported by the
US Department of Energy
and National Science Foundation, the
Natural Sciences and Engineering Research Council (Canada),
Institute of High Energy Physics (China), the
Commissariat \`a l'Energie Atomique and
Institut National de Physique Nucl\'eaire et de Physique des Particules
(France), the
Bundesministerium f\"ur Bildung und Forschung and
Deutsche Forschungsgemeinschaft
(Germany), the
Istituto Nazionale di Fisica Nucleare (Italy),
the Foundation for Fundamental Research on Matter (The Netherlands),
the Research Council of Norway, the
Ministry of Science and Technology of the Russian Federation, and the
Particle Physics and Astronomy Research Council (United Kingdom). 
Individuals have received support from 
CONACyT (Mexico),
the A. P. Sloan Foundation, 
the Research Corporation,
and the Alexander von Humboldt Foundation.